# Charge density waves and enhanced superconductivity in the electron-hole gas:

## a plausible, simple, physically intuitive model

**Carl A. Kukkonen**
33841 Mercator Isle, Dana Point, California 92629, USA
kukkonen@cox.net

## ABSTRACT

A neutral degenerate plasma of equal numbers of electrons and positively charged fermions (holes) — the electron-hole gas — is studied using a simple, physically motivated model. Each species is treated as an independent Fermi gas in a uniform background of opposite charge, coupled only by their mutual Coulomb attraction. The model has two parameters: the density $r_s$ and the mass ratio M/m. The following two simplifying assumptions are made and examined but not proven: the missing electron-hole correlation energy is approximately constant with density and therefore does not contribute to the pressure or bulk modulus, the known local field factors of the electron gas can be used, and the electron-hole contribution to the local field factors is zero. With these assumptions the phase diagram, instabilities, and effective interactions can be calculated for all mass ratios and densities. These are simple formulae and all quantitative results presented in this paper can be calculated on a laptop.

The key new physics in the electron-hole gas is the additional screening provided by the mobile holes. A single function Δ appears in the denominator of all effective interactions and response functions. In addition to the compressibility instability when $\Delta = 0$ at $q = 0$, near the zeros of Δ at finite q three phenomena are simultaneously enhanced: a charge density wave instability (for $M/m \geq 4.97$), a large $T^2$ electrical resistivity from electron-hole scattering, and an attractive term in the effective electron-electron interaction that is the purely electronic analog of BCS electron-phonon coupling. Crudely estimated superconducting transition temperatures approach room temperature, reaching approximately 275 K at $M/m = 9$.

No claim is made that this model applies to any specific material. The aim is to provoke scrutiny of the assumptions, calculation of the unknown local field factors, and investigation of whether this minimal two-carrier framework maps onto real systems.

## I. INTRODUCTION

Three of the most intensely debated phenomena in condensed matter physics — high-temperature superconductivity, charge density waves, and an anomalous $T^2$ electrical resistivity — have a peculiar habit of appearing together. In the cuprates [23], iron pnictides, and other families of unconventional superconductors, charge density wave order and superconductivity compete and coexist in adjacent regions of the phase diagram, while the normal state above the superconducting transition often exhibits a $T^2$ resistivity that is far larger than expected from conventional electron-electron scattering. Whether these three phenomena share a common origin or merely coexist by coincidence is a central unsolved question. This paper shows that a remarkably simple model — a neutral gas of electrons and mobile positive fermions interacting purely through the Coulomb interaction, with no phonons, no lattice, and no adjustable parameters beyond the density and the mass ratio — contains all three phenomena simultaneously, arising from a single underlying mechanism.

The individual components of this gas are well understood and entirely featureless. The uniform electron gas — fermions of mass $m$ and charge −e in a rigid neutralizing background — is characterized by a single density parameter $r_s$, and after a century of study its properties at $T = 0$ are thoroughly catalogued [1]. It supports no charge density waves and no superconductivity (see Ref. [22] for a very low temperature exception). The uniform positive fermion gas is the same problem with the mass replaced by $M$; because the Schrödinger equation depends only on the square of the charge, all electron-gas results carry over with a mass rescaling. Alone, the hole gas is equally stable and uninteresting.

The two-component electron-hole gas is qualitatively different from either of its constituents, for two reasons. First, both species are mobile: a positive disturbance attracts electrons and repels holes, setting both in motion simultaneously. Second, electrons and holes attract each other, so the screening response of each species modifies that of the other. This mutual screening — absent in any single-component system — is the essential new many-body effect. It introduces a second parameter with no single-component analog: the mass ratio $M/m$. The formal linear response theory of the two-component gas was worked out by Vashishta, Bhattacharyya and Singwi in 1974 [4].

The entire singular behavior of the electron-hole gas is captured by a single real function Δ, which depends on wave vector $q$, density $r_s$, and mass ratio $M/m$. To see why, consider the response to an external positive test charge with potential $V_{ext}$. In the pure electron gas the screened interaction seen by another electron is

$$V_{et} = V_{ext} / \varepsilon_{et} \qquad (1)$$

where the dielectric function $\varepsilon_{et}$ is always positive: no instabilities occur. In the electron-hole gas the corresponding expression becomes

$$V_{et} = V_{ext} \cdot H / \Delta \qquad (2)$$

where $H$ is a well-behaved positive function. All singular behavior resides in Δ. When $\Delta \to 0$ at $q = 0$, the bulk modulus of the system vanishes: a compressibility instability signaling that the uniform gas is no longer the lowest-energy state. When $\Delta \to 0$ at finite $q$, the response to a density perturbation at that wave vector diverges: a charge density wave. A 1985 investigation [5] examined this problem but employed the Hubbard approximation for the wave-vector-dependent local field factor, which underestimates the local field at intermediate $q$, and did not find charge density waves. With the accurate, quantum Monte Carlo-consistent local field factors now available, both instabilities are found and the phase diagram can be mapped quantitatively.

The same quantity Δ governs the effective electron-electron interaction, but its role there is more striking still. The effective interaction between two electrons is richer than the test-charge response because a real electron has spin, finite mass, and is indistinguishable from other electrons of the same spin. These requirements, first worked out from physical arguments in 1979 [6] and confirmed diagrammatically in 1985 [7, 8], yield

$$V_{ee}^{eff}(\sigma_1, \sigma_2, q, \omega) = V_0(q,\omega) - J(q,\omega)\, \sigma_1 \cdot \sigma_2 \qquad (3)$$

where $\sigma_1$, $\sigma_2$ are the Pauli spin matrices, $V_0 > 0$ is repulsive, and $J < 0$ makes parallel-spin scattering slightly less repulsive than antiparallel. In the pure electron gas this interaction is always repulsive. Earlier work [22] showed that the electron gas alone possesses a pairing instability through higher angular momentum channels, but at transition temperatures so low as to be physically irrelevant. In the electron-hole gas, the mutual screening between the two species generates an attractive mechanism many orders of magnitude larger.

In the electron-hole gas, the mobile holes open a new screening channel. Kukkonen and Overhauser [6] showed that any elastically deformable charged medium generates an additional attractive contribution to the effective electron-electron interaction but left the result in general form. Vignale and Singwi [5] derived the explicit expression for the electron-hole gas:

$$V_{ee}^{eff}(\sigma_1, \sigma_2, q, \omega) = V_0(q,\omega) - J(q,\omega)\, \sigma_1 \cdot \sigma_2 - U^{eh} \qquad (4)$$

where the first two terms are unchanged from the single-component case, and the new third term is attractive and proportional to 1/Δ.

The physical meaning of this term is transparent when placed alongside conventional BCS theory. In BCS, the positive ions are classical, stationary particles on a lattice whose Coulomb interaction with the electrons is conveyed through lattice vibrations — phonons. The resulting phonon-mediated attraction between electrons is what drives Cooper pairing. The term $-U^{eh}$ in Eq. (4) plays an analogous role: it is the electron-phonon attraction of BCS, but generalized to a medium of fully mobile, quantum-mechanical positive fermions. The mass ratio $M/m$ is the parameter that encodes this generalization. In the limit $M/m \to \infty$ the holes become infinitely heavy, their kinetic energy vanishes, and one recovers the classical deformable-medium limit that BCS implicitly assumes. For finite $M/m$ the holes are genuinely quantum mechanical, and near the instabilities where $\Delta \to 0$ the attractive term diverges. There are no phonons in this model; the attraction is a purely electronic, Coulomb effect.

The normal-state transport is governed by the same Δ. In the electron-hole gas, electron-electron and hole-hole collisions conserve the momentum of each species and therefore do not contribute to the electrical resistivity; only electron-hole scattering, which transfers momentum between the two species, generates resistance. Kukkonen and Maldague showed in the 1970s [9, 10] that the resulting resistivity varies as $T^2$, with a coefficient proportional to $1/\Delta^2$. Near an instability this coefficient diverges, producing a large $T^2$ resistivity in the normal state. All three scattering processes contribute to the thermal resistivity, and the ratio of the electrical resistivity to the thermal resistivity could be interesting.

The central observation of this paper is that the single function $\Delta$ simultaneously controls all three phenomena. As $\Delta \rightarrow 0$: the attractive interaction $-U^{eh}$ (□ $1/\Delta$) diverges and Cooper pairing is strongly enhanced; the $T^2$ resistivity coefficient $1/\Delta^2$ diverges and the normal state becomes anomalously resistive; and the response to density perturbations diverges, driving either a compressibility instability or a charge density wave. These are not three separate effects — they are three faces of the same instability of the two-component Coulomb gas. A purely electronic mechanism, with no phonons and no spin fluctuations, produces phenomenology that echoes motifs seen in the strange-metal and pseudogap regimes of the high-temperature superconductors.

This coincidence does not prove that the electron-hole gas model underlies high-temperature superconductivity in any real material. The present work makes no such claim. What it does establish is that within the approximations stated, a model of irreducible simplicity — fully specified by two parameters, requiring no material-specific assumptions, and containing no free variables to tune — predicts the simultaneous appearance of charge density waves, enhanced superconductivity, and large $T^2$ resistivity as a consequence of a single function approaching zero. Whether this minimal framework can be mapped onto real semimetals, compensated metals, or the electron-hole plasma in correlated oxides is an open and, we suggest, important question. The following sections provide the explicit expressions for $\Delta$ and the effective interactions in terms of the local field factors, which are the only material input the theory requires.

## II. LINEAR RESPONSE THEORY AND LOCAL FIELD FACTORS

A brief overview of linear response theory and local field factors is useful to introduce the concepts and notation. A more detailed derivation is given in Refs. [2, 3], the original work is in Ref. [4], and a thorough discussion appears in the textbook [1]. Throughout, subscript 1 refers to holes and subscript 2 refers to electrons.

A small positive test charge perturbation is introduced into the electron-hole gas with bare Coulomb potential $V_{ext} = v\rho_{ext}$, where the Coulomb potential is $v = 4\pi/q^2$. The test charge attracts electrons, inducing a local density increase $\Delta n_2$, and repels holes, creating a local density decrease $\Delta n_1$. A specific electron in the gas has a bare Coulomb interaction with both induced densities. It also has an exchange interaction with the induced electron density $\Delta n_2$ — since half the electrons have the same spin in an unpolarized gas — and a correlation interaction with $\Delta n_2$ arising from their mutual Coulomb repulsion. Together these give rise to the electron-electron local field factor $G_{22}$. The specific electron also has a correlation interaction with the induced hole density $\Delta n_1$ due to their mutual Coulomb attraction, giving rise to the off-diagonal local field factor $G_{21}$.

The effective potentials felt by an electron and a hole due to the external test charge are

$$V_{2t} = V_{ext} - v(1-G_{21})\, \Delta n_1 + v(1-G_{22})\, \Delta n_2 \qquad (5)$$

$$V_{1t} = V_{ext} + v(1-G_{11})\, \Delta n_1 + v(1-G_{12})\, \Delta n_2 \qquad (6)$$

First-order perturbation theory relates the induced densities to the effective potentials through the free-particle (Lindhard) response functions $\Pi^0$: $\Delta n_1 = -\Pi^0_1 V_{1t}$ and $\Delta n_2 = -\Pi^0_2 V_{2t}$. By symmetry $G_{12} = G_{21}$. Solving these two coupled equations yields the effective potentials in terms of the local field factors:

$$V_{1t} = (1 + (G_{22} - G_{12})\, v\Pi^0_2)\, V_{ext} / \Delta \qquad (7)$$

$$V_{2t} = (1 + (G_{11} - G_{12})\, v\Pi^0_1)\, V_{ext} / \Delta \qquad (8)$$

$$\Delta = (1 + (1-G_{11})\, v\Pi^0_1)\, (1 + (1-G_{22})\, v\Pi^0_2) - (v\, (1-G_{12}))^2\, \Pi^0_1\, \Pi^0_2 \qquad (9)$$

These are precisely the equations derived in Ref. [4] for a positive charge perturbation, where the perturbation has opposite signs for electrons and holes. Note that the last term in Eq. (9) is negative, so $\Delta$ can become small or zero if that term becomes comparable to the first. This is what motivates further investigation.

For the effective electron-electron interaction, the expressions for $V_0$, $J$, and $U^{eh}$ in Eq. (4) in terms of the local field factors, derived in Refs. [5–8], are:

$$V_0 = v\, (1 + (1-G_{22})\, G_{22}\, v\Pi^0_2) / (1 + (1-G_{22})\, v\Pi^0_2) \qquad (10)$$

$$J = -v\, G^2_{22M}\, v\Pi^0_2 / (1 - G_{22M}\, v\Pi^0_2) \qquad (11)$$

$$U^{eh} = v(v\Pi^0_1\, (1-G_{12}))^2 / (\Delta\, (1 + (1-G_{22})\, v\Pi^0_2)) \qquad (12)$$

Equation (12) defines $U^{eh}$ to include the factor $1/\Delta$ explicitly. Since the attractive term in Eq. (4) is $-U^{eh}$ □ $1/\Delta$, proximity to an instability $\Delta \rightarrow 0$ directly enhances the electron-electron attraction.

These equations follow the notation of Refs. [2, 3]. The local field factor $G_{22M}$ governs the magnetic (spin) response of the electron gas. The advantage of writing the theory in terms of local field factors is that it separates what is known exactly — the structure of the response functions — from what must be calculated: the local field factors themselves.

The introduction of the local field factors also makes explicit where the additional electron-hole physics enters. The term $U^{eh}$ in Eq. (12) contains $G_{12}$, the off-diagonal local field factor encoding electron-hole correlations. The off-diagonal term $G_{12}$ is unknown for the electron-hole gas and is one of the key quantities whose calculation this paper aims to motivate.

It should be noted that the argument that the electron-hole correlation energy is approximately independent of $r_s$ justifies its omission from the bulk modulus and equilibrium density, but does not by itself justify setting $G_{12} = 0$. The latter is a separate and additional approximation, made on physical grounds, that the off-diagonal local field factor encoding electron-hole correlations at finite $q$ is small compared to the diagonal terms. Both assumptions are stated explicitly in Section III and represent the two principal uncertainties of the model.

## III. LOCAL FIELD FACTORS

The local field factors of the electron-hole gas are not known. They encode and represent the complicated, nonlocal many-body effects in the electron-hole gas. A goal of this short paper is to encourage calculations of these local field factors.

The existence of $\Delta$ in the denominator of the effective interactions in the electron-hole gas was known for 40–50 years, but largely ignored. This was likely because it is difficult to calculate. Perhaps it was assumed that it was a rather dull function like the electron gas dielectric function.

The contribution of this paper is to make an obvious conjecture that, in the first approximation, the local field factors for the electron-hole gas are the same as for the electron gas in a uniform background and the hole gas in a uniform background, and that the off-diagonal term $G_{12} = 0$. This allows exploration of $\Delta$ and the effective interactions as a function of $r_s$ and mass ratio $M/m$.

The physical picture used here is that the positive fermions or holes are the uniform background for the electrons and vice versa. In the usual electron gas, the lattice is approximated by a uniform background and it is assumed that the electronic properties are governed by the Fermi gas of electrons. Here there is also a Fermi gas of holes. These two systems are coupled by the Coulomb attraction between the electrons and holes.

For the single-component electron gas, the electron density is uniform as long as the background is uniform. The same is true for a hole gas. Therefore the assumption made here seems reasonable as long as both species remain uniform. If $\Delta$ becomes zero, the density is no longer uniform and the assumption breaks down. This is a phase transition.

The ground state energy of the electron-hole gas in this approximation is simply the sum of the two independent gases in a uniform background. What is included is the kinetic energy, exchange energy, and electron-electron and hole-hole correlation energies. The electron-hole correlation energy is ignored. In Ref. [2] many previous calculations of the electron-hole correlation energy are examined and it is found that for a given mass ratio $M/m$ it is a weak function of $r_s$. This means that the missing energy does not strongly influence the $r_s$ of the energy minimum (pressure $P = 0$), which is determined by the first derivative of the ground state energy with respect to volume, or that of the compressibility instability where the total bulk modulus $B$ equals zero. Note that calculations of the ground state energy of the electron-hole gas that include the electron-hole correlation energy are very close to the energy of an electron-hole exciton, and that the electron-hole gas may be metastable or unstable.

It should be noted that the argument that the electron-hole correlation energy is approximately independent of $r_s$ justifies its omission from the bulk modulus and equilibrium density, but does not by itself justify setting $G_{12} = 0$. The latter is a separate and additional approximation, made on physical grounds, that the off-diagonal local field factor encoding electron-hole correlations at finite $q$ is small compared to the diagonal terms. Both assumptions represent the two principal uncertainties of the model.

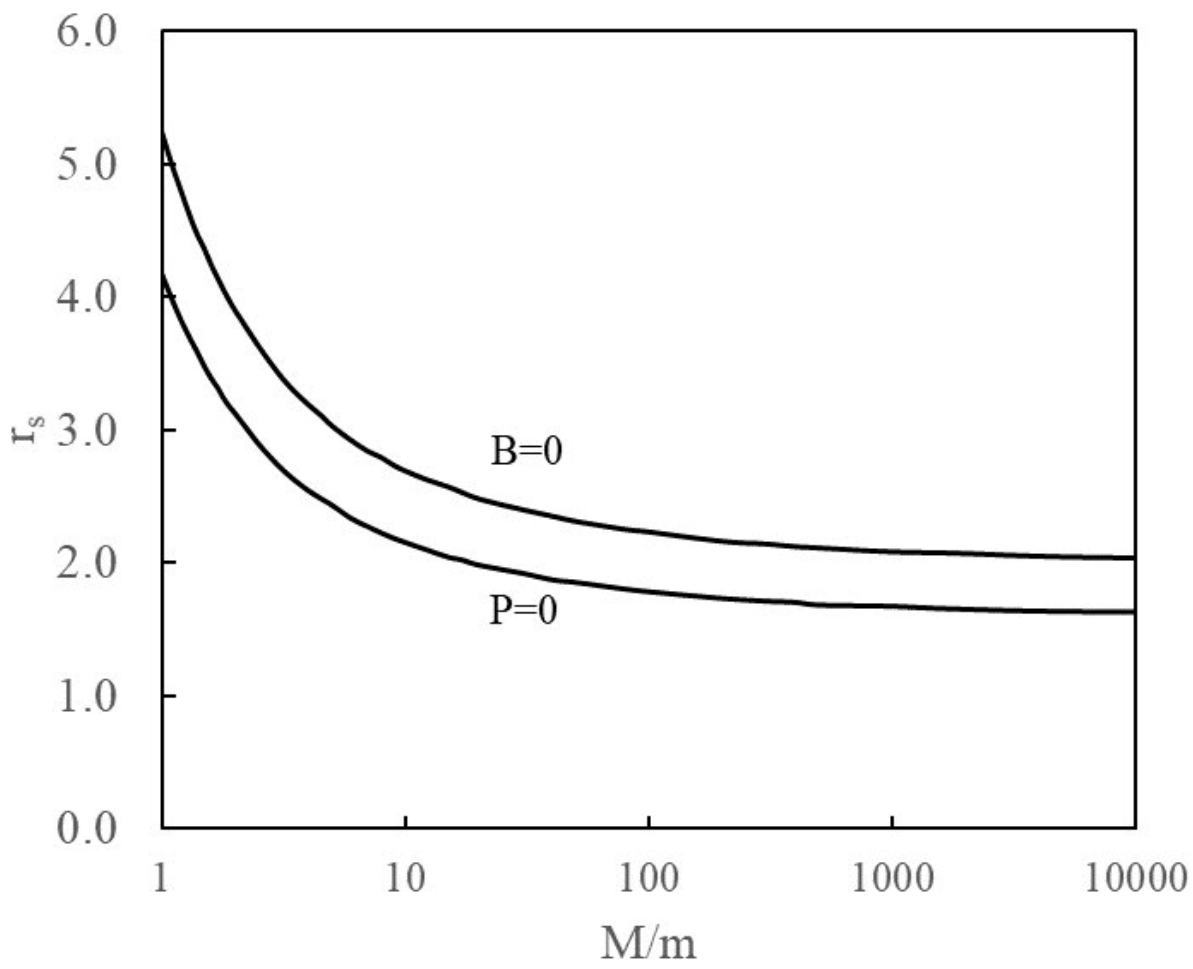

FIG. 1. Equilibrium density $r_s$ where the electron-hole gas has an energy minimum and the total pressure P = 0. The electron-hole gas becomes thermodynamically unstable when the bulk modulus B = 0. This point is known as the compressibility instability. Note that the $r_s$ of the B = 0 point is closely 1.25 that of the equilibrium $r_s$ where P = 0.

Figure 1 shows that the electron-hole gas has an equilibrium $r_s$ that varies from 4.19 (the same as the electron gas in a uniform background) when $M/m$ = 1, and is asymptotic to $r_s$ = 1.63 for large $M/m$. The reason for the asymptotic behavior is that the kinetic energy of the holes becomes negligible for large mass, while the exchange energy is independent of mass and the correlation energy becomes independent of mass for large $M$. If there were no other factors, the electron-hole gas would sit at the equilibrium $r_s$.

At the equilibrium $r_s$, the electron gas pressure is positive and the hole gas pressure is negative. At the compressibility instability where the bulk modulus $B$ = 0, the electron bulk modulus is positive and the hole bulk modulus is negative. The lighter electrons stabilize the system, in contrast to the electron gas problem where it is assumed that the rigid background provides the stabilizing force controlling the $r_s$ of the system. For the electron-hole gas with no other external factors, it will always sit at the equilibrium $r_s$. In order to map out the phase diagram, $r_s$ is treated as an independent variable along with the mass ratio $M/m$.

The most important result from the quantum Monte Carlo calculations of the electron gas is that the density local field factor goes as $q^2$, not only at very small $q$ as required by the compressibility sum rule, but continuing this quadratic behavior up to approximately $2k_F$:

$$G_{22} = (1 - \kappa_0/\kappa)(q/q_{TF})^2 = [(1 - B/B_0)(3\pi^2/16)^{2/3}/r_s](q/k_F)^2$$

The Fermi wave vector is $k_F$. The compressibility is $\kappa$ and the subscript 0 signifies the compressibility of the noninteracting electron gas. The bulk modulus $B$ is the inverse of the compressibility. This continued $q^2$ behavior is responsible for the charge density waves that occur between 0.83 and $2k_F$.

With these assumptions, $\Delta(q)$ is calculated as a function of wave vector for any given mass ratio $M/m$ and density $r_s$. The effective interactions are proportional to $1/(q^2\Delta)$, which makes it natural to examine the quantity $\Delta\cdot(q/k_F)^2$; when this quantity is zero the electron-hole system is unstable.

For mass ratios below $M/m = 4.97$, the only zeros of $\Delta\cdot(q/k_F)^2$ occur at $q = 0$ (compressibility instability). At $M/m = 4.97$, two zeros occur simultaneously at $q = 0$ and $q = 0.83k_F$ at $r_s = 3.036219$. For mass ratios $M/m > 4.97$, finite-$q$ charge density wave instabilities occur at $r_s$ less than the $r_s$ of the compressibility instability.

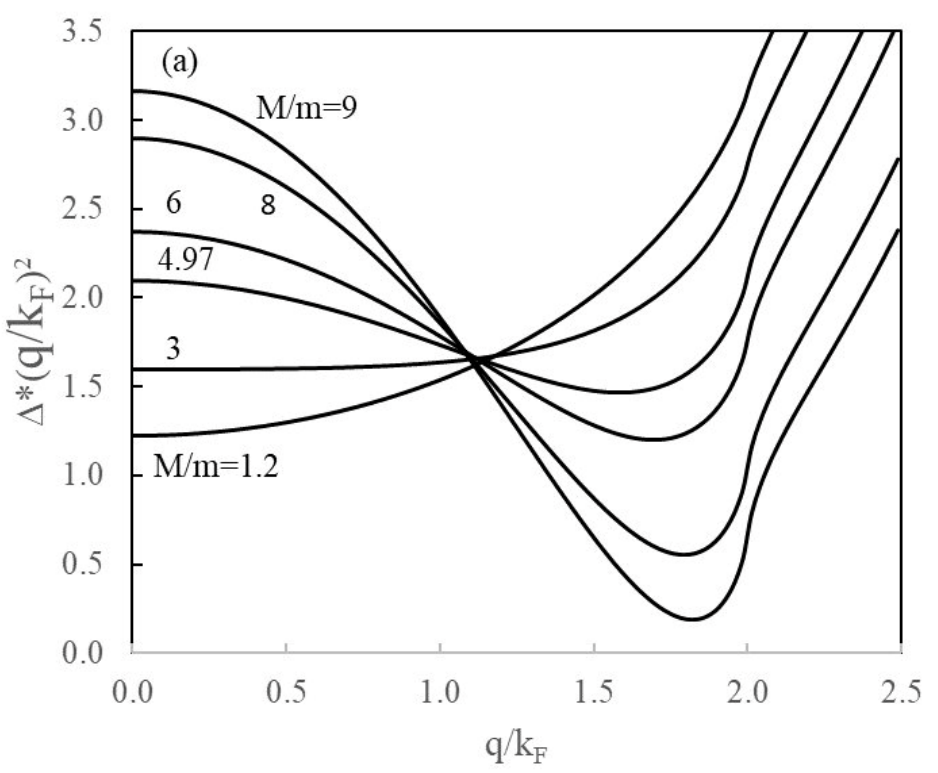

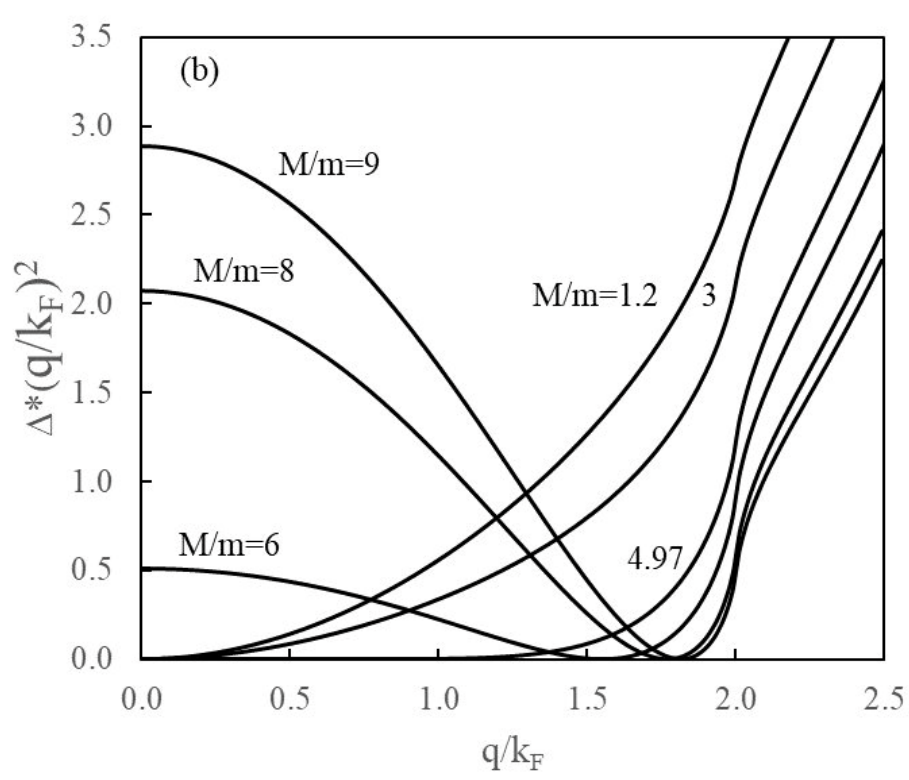

FIG. 2. The electron-hole gas in this simple model has a unique minimum of its ground state energy at each mass ratio M/m; this is the equilibrium density. (a) $\Delta\cdot(q/k^M)^2$ at the equilibrium density for representative mass ratios. At small mass ratios the curves increase monotonically; for larger mass ratios the decrease at large q reflects the incipient charge density wave. (b) $\Delta\cdot(q/k^M)^2$ evaluated at the $r_s$ where the quantity first reaches zero, obtained by increasing $r_s$ from its equilibrium value. At small mass ratios the instability is at q = 0 (compressibility instability); at larger mass ratios it is at finite q (charge density wave).

The special case of $M/m = 4.97$ deserves attention. In Figs. 2(b) and 3(a), $\Delta\cdot(q/k_F)^2$ appears to remain close to zero until about $q/k_F = 1$. A 100× magnification shown in Fig. 3 reveals two simultaneous zeros. This is the triple point where the normal state, the compressibility instability, and the charge density wave coexist.

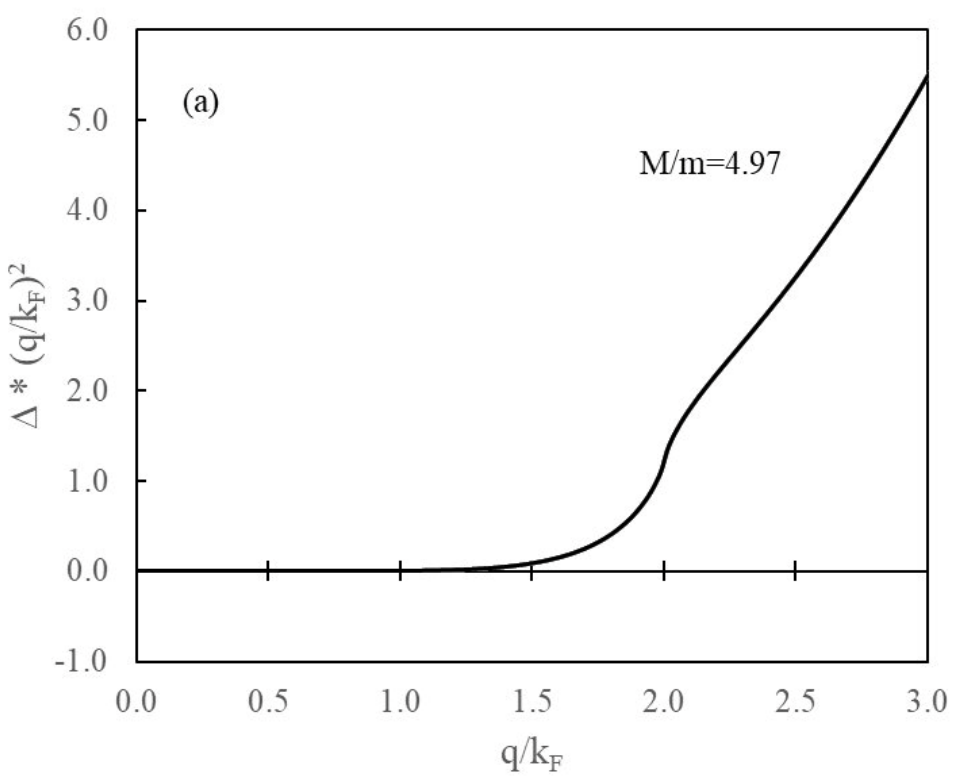

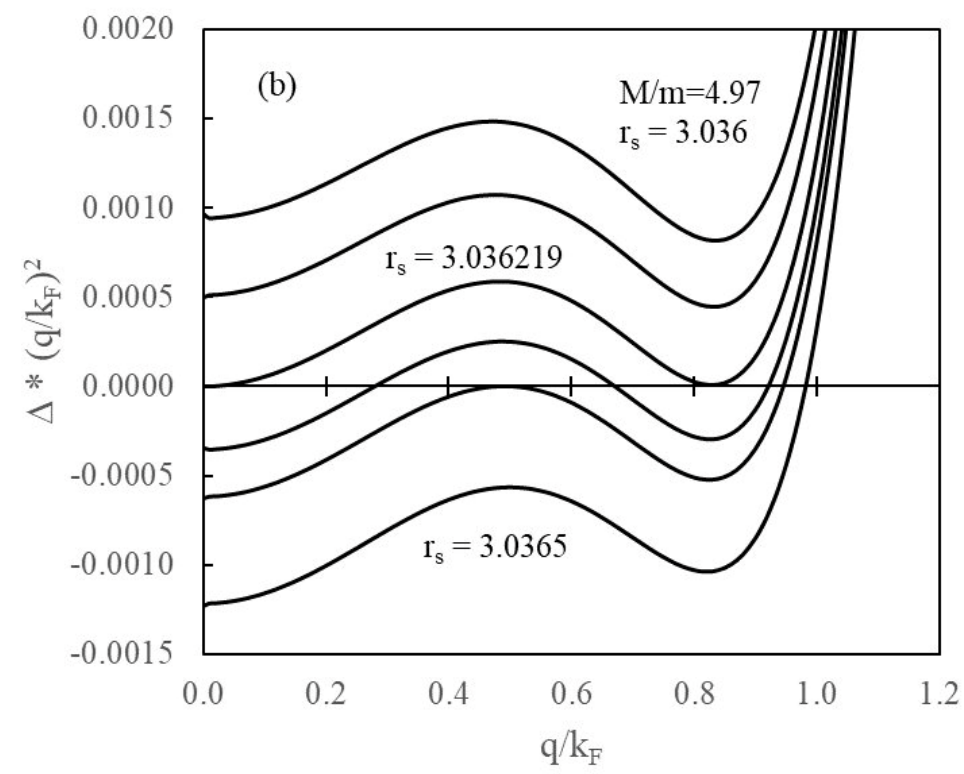

FIG. 3. The triple point at M/m = 4.97, $r_s = 3.036219$. (a) $\Delta\cdot(q/k^M)^2$ at the triple point density, which appears close to zero over a wide range of q. (b) 100× magnification showing the two simultaneous zeros at q = 0 and $q = 0.83k^M$, confirming the coexistence of the compressibility instability and the charge density wave at a single density.

Near the finite-$q$ instability, the point where $\Delta\cdot(q/k_F)^2 = 0$ is very sensitive to both the density $r_s$ and the mass ratio $M/m$. This is shown clearly in Fig. 4.

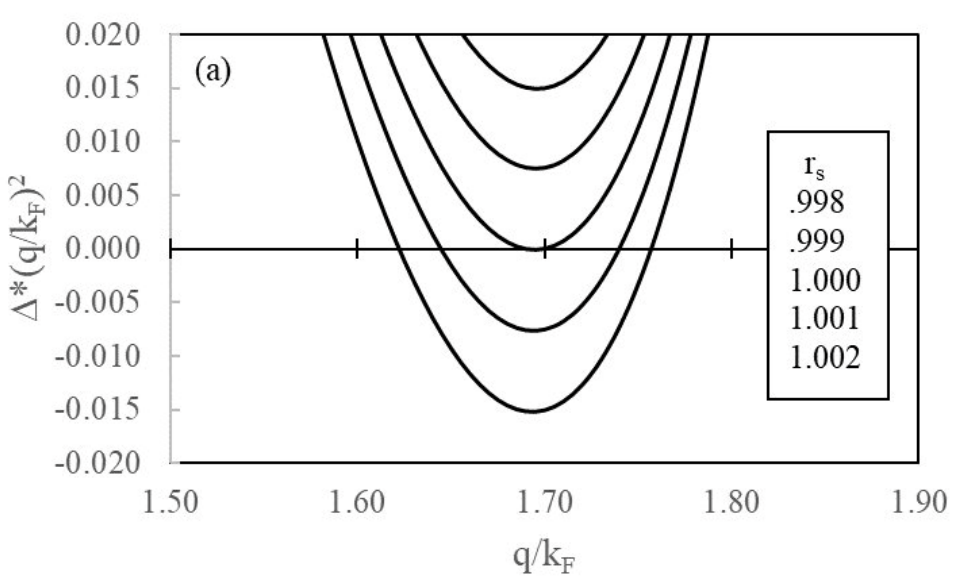

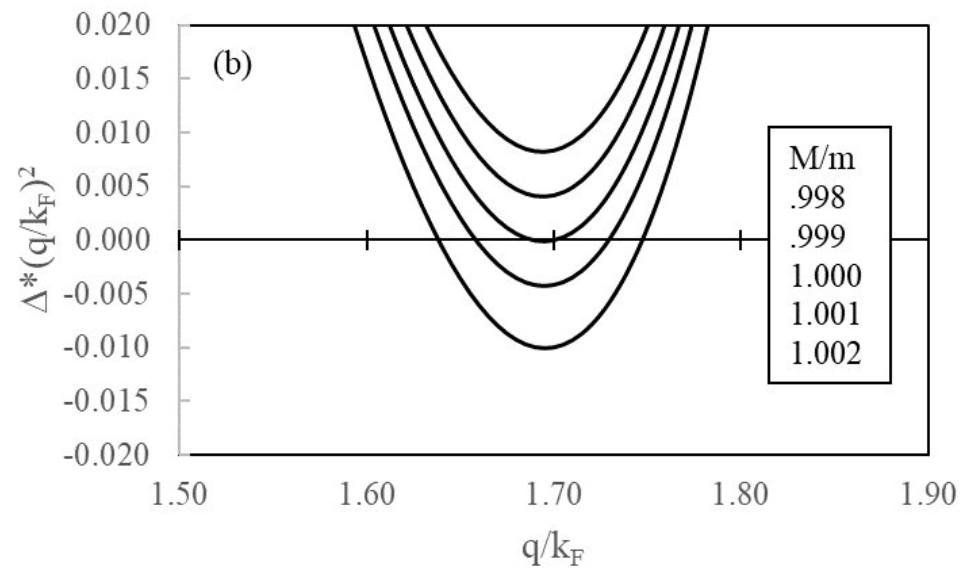

FIG. 4. High sensitivity of the charge density wave instability to small changes in $r_s$ and M/m, illustrated for M/m = 7 (charge density wave at $r_s = 2.60779$, $q/k^M = 1.69$). (a) $r_s$ varied by the amounts indicated. (b) M/m varied by the amounts indicated. A change of 0.1% in either variable is sufficient to trigger or suppress the charge density wave.

Although there is no attempt here to map this model onto a real system, if such a mapping were made the resulting electron-hole gas would have other influences besides the mutual Coulomb interaction. Effective masses and the dielectric function of the ionic background would need to be considered.

Consider a system with $\Delta\cdot(q/k_F)^2$ slightly greater than zero (no actual charge density wave). Figure 4(b) shows that increasing the mass ratio $M/m$ by as little as 0.1% would trigger the charge density wave. This might correspond to a distortion of the lattice, reminiscent of twisting in graphene. A larger hole mass means a flatter band.

A simple physical picture helps explain why the mass ratio $M/m$ controls the strength of the instabilities. A heavier hole has lower Fermi pressure and a larger static density response at the same $r_s$ — the hole gas is more easily polarized. As *M/m* increases, the combined charge response of the two-component system softens, finite-*q* charge modes become progressively easier to trigger, and the induced attraction in the electron-electron channel grows. The electrons, with higher Fermi pressure, stabilize the system, while the heavier holes supply an increasingly deformable positive medium. This is why the instabilities and their associated enhancements become more pronounced as *M/m* increases.

After completing the work reported in Refs. [2, 3], a closely related paper was discovered: "Charge density waves in a quantum plasma" [14]. Those authors calculated the same phase diagram using density functional theory within the local spin density approximation, also neglecting the additional correlation effects between electrons and positive fermions. The two phase diagrams are nearly identical. This close agreement validates the simple model. Reference [14] provides a good discussion and suggests connections to possible exotic phase transitions, new intermediate phases, and smectic liquid crystals. Those authors emphasized that the $q^2$ behavior of the local field factor near $q = 1–2k_F$ is the source of the charge density waves. They did not calculate the effective interactions or recognize the important influence of the instabilities even in the normal phase.

The early papers [4, 5] used the Hubbard approximation for the wave vector dependence of the local field factors. The Hubbard approximation goes to a constant less than one at large *q* and never becomes large enough to trigger the charge density wave. This is the crucial distinction introduced by the quantum Monte Carlo results.

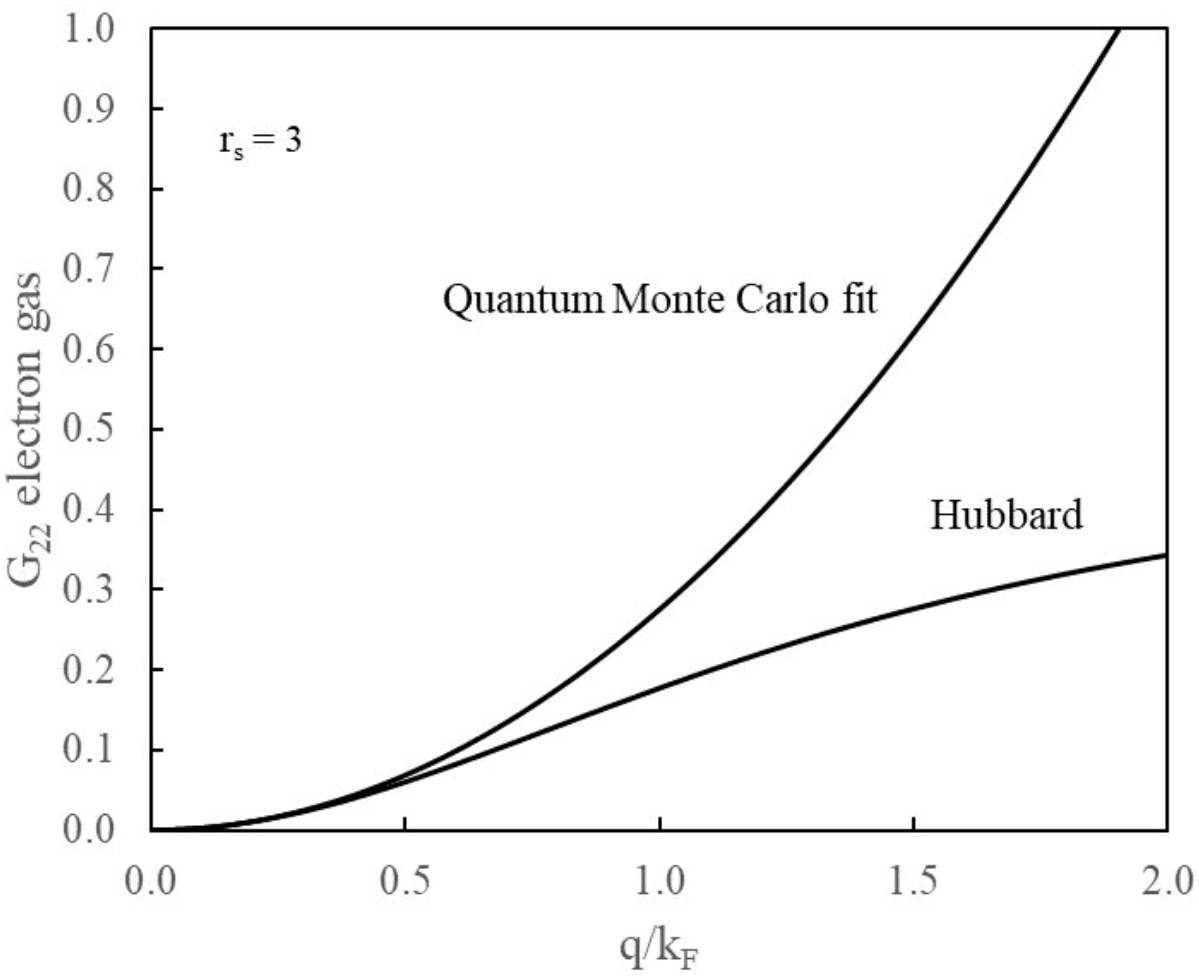

FIG. 5. Local field factor $G_{22}$ for the electron gas in a uniform rigid neutralizing positive background. The Quantum Monte Carlo fit is a simple quadratic that fits the actual quantum Monte Carlo local field factor quite well [13] and is equivalent to the local density approximation in density functional theory. The earlier Hubbard approximation was used previously to represent the wave vector dependence of the local field factor. Both agree at small q because they both satisfy the compressibility sum rule. Using the Hubbard approximation, there are no charge density waves.

The main message of this section is that near an instability, $\Delta\cdot(q/k_F)^2$ is very sensitive to both $r_s$ and $M/m$, and even reasonably far from an instability it is strongly influenced by the incipient instabilities. Although no attempt is made here to map this model onto any real material, any experimental method that varies $r_s$ or $M/m$ could produce dramatic effects. An easy way to manipulate $r_s$ is to apply pressure, which decreases $r_s$. Twisting or applied torque may affect the effective mass ratio.

## IV. ELECTRON-HOLE INTERACTION

The general equations governing interactions were derived in the supplemental material to Ref. [2] using the same approach as in Eqs. (5–9). The key is to determine the external potential perturbing the system. To calculate the electron-hole effective interaction, the perturbation is taken to be a positive fermion with spin up. Since this positive fermion is identical to other positive fermions with spin up, there is an exchange and correlation interaction with them. There is a correlation interaction with spin down positive fermions, and a correlation interaction with electrons. The positive fermion has finite mass and has a correlation interaction with both species. It is important to note that the "external" potential felt by holes is different from that of the electrons. The attractive effective interaction between an electron and hole is simply

$$V_{21}^{\mathrm{eff}} = -v(1 - G_{12}) / \Delta \qquad (13)$$

In the simple model of this paper I take $G_{12} = 0$. A sensitivity analysis is done in Ref. [2] and the wave vector dependence is $q^2$ so that it has no effect at small $q$ and modifies the quantitative results somewhat but does not affect the overall conclusions. A goal of this paper is to encourage calculations of $G_{12}$.

It is instructive to compare this to the RPA where all local field factors equal zero:

$$V_{21}^{\mathrm{eff,RPA}} = -v / (1 + v\Pi^0{}_1 + v\Pi^0{}_2) \qquad (14)$$

and the effective interaction if only the electrons did the screening:

$$V_{21}^{\mathrm{eff}} = v / \varepsilon_{et} = v / (1 + (1-G_{22})\, v\Pi^0{}_2) \qquad (15)$$

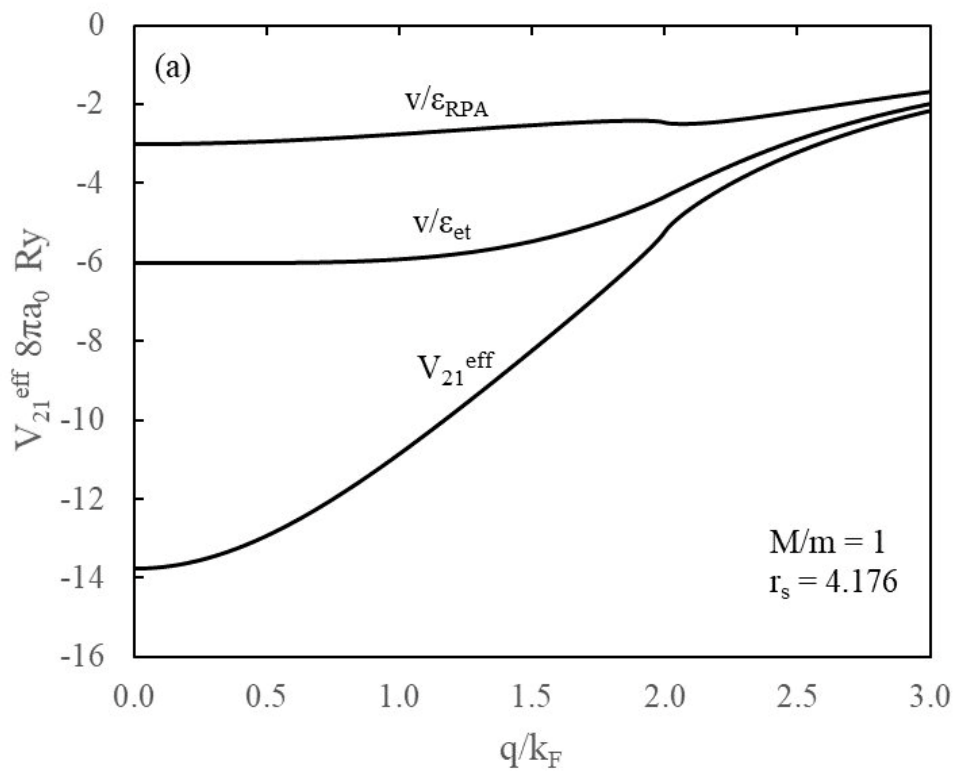

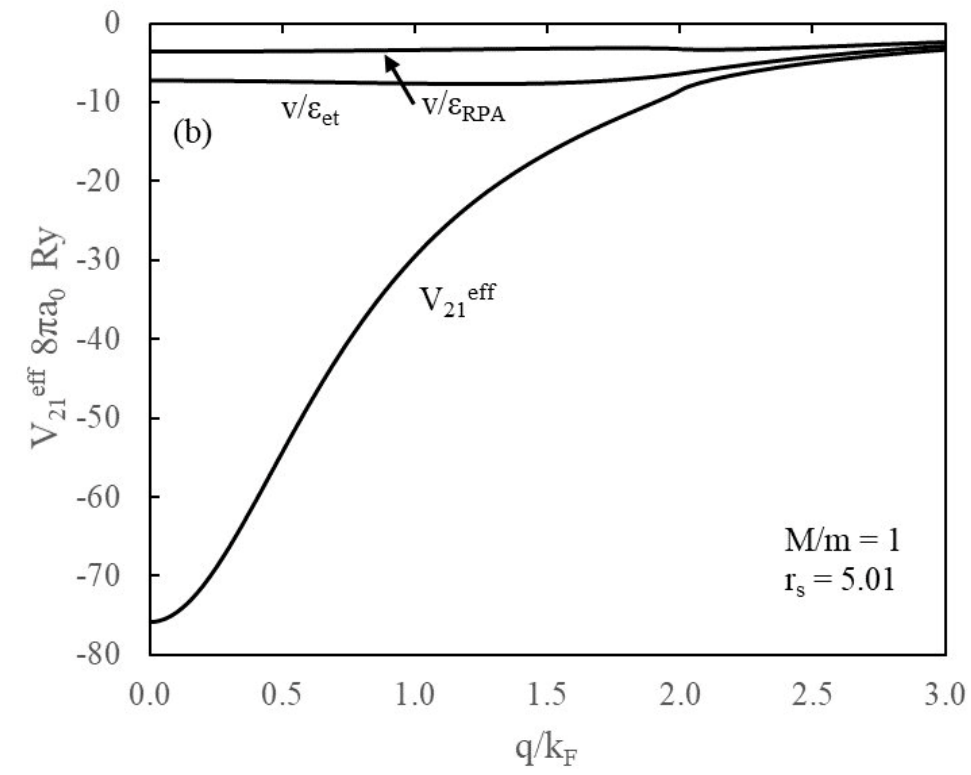

FIG. 6. The effective attractive interaction between electrons and positive fermions (holes) as a function of wave vector at mass ratio M/m = 1 for which the equilibrium density is $r_s$ = 4.176. The effective interaction from the simple model is compared to the RPA and the Coulomb potential divided by the electron-test charge dielectric function. Fig. 6a shows the effective interaction at equilibrium density $r_s$ = 4.176, and Fig. 6b shows it at $r_s$ = 5.01 which is closer to the compressibility instability at $r_s$ = 5.248.

For *M/m* = 1, the effective electron-hole interaction is most attractive at $q$ = 0. Figure 6a shows that the RPA and electron-only screening are poor approximations. Figure 6b shows the same comparison at $r_s$ = 5.01, closer to the instability, revealing the large effect of the incipient compressibility instability. $V_{21}^{\text{eff}}$ is very large and attractive near $q$ = 0; this is beyond linear response theory and another approach is needed. The attractive interaction suggests a possible bound state or exciton.

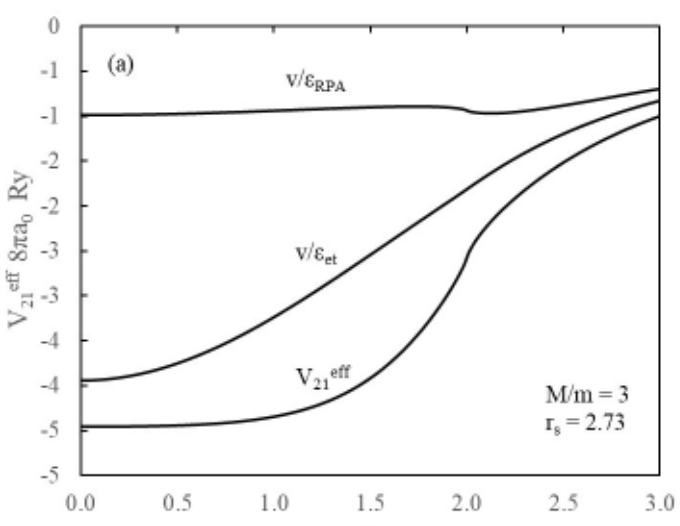

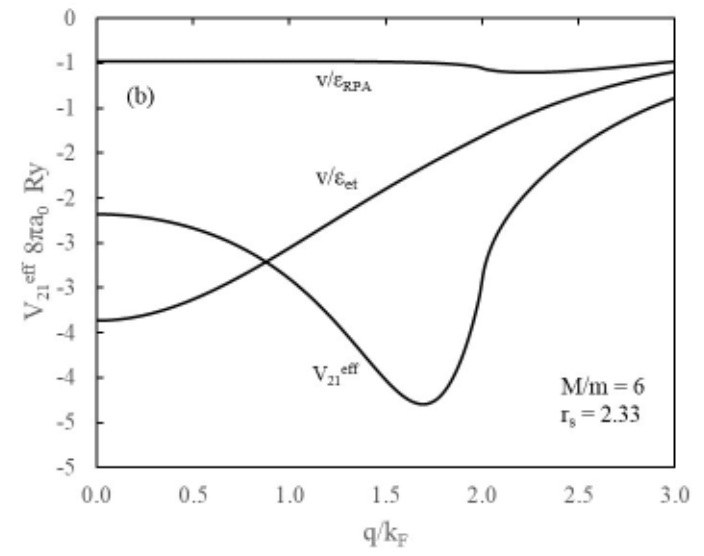

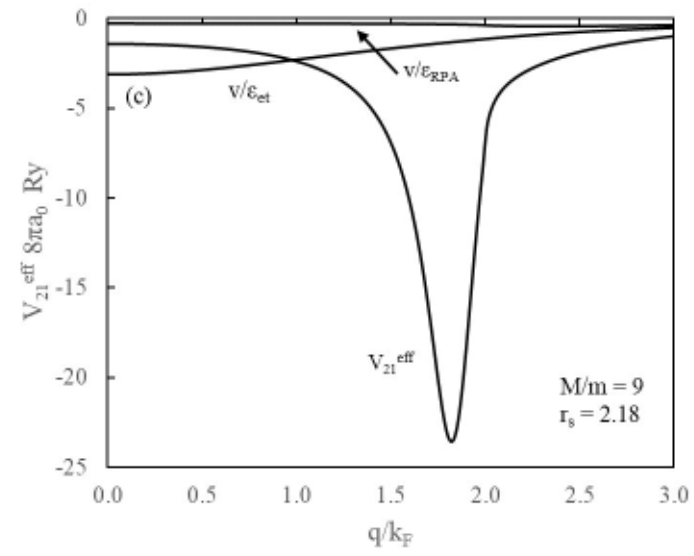

FIG. 7. The effective attractive electron-positive fermion (hole) interaction versus wave vector plotted at the equilibrium $r_s$ for each mass ratio. (a) M/m = 3, $r_s$ = 2.73; (b) M/m = 6, $r_s$ = 2.33; (c) M/m = 9, $r_s$ = 2.18. The different interactions are described in the caption for Fig. 6.

Even at mass ratio *M/m* = 3, $V_{21}^{\text{eff}}$ is flatter and more attractive at finite $q$. At *M/m* = 6 there is a pronounced dip near $q/k_F$ = 1.8 which is the wave vector of the incipient charge density wave. At *M/m* = 9 the equilibrium density is close to the critical density for the charge density wave and the effect is huge. Large $q$ scattering is more effective at degrading the electrical current.

Figures 6 and 7 show that the incipient instabilities significantly increase the effective electron-hole attraction, and that the RPA and the electron-test charge dielectric function are poor approximations to the full screened interaction.

## V. ELECTRON-ELECTRON INTERACTION

The objective is to calculate the effective electron-electron interaction needed for superconductivity and for normal state electron-electron scattering that contributes to transport properties. The following is a summary of the physics considerations in deriving the interaction which is done in detail in Refs. [2, 5, 6, 8].

Consider two particular electrons interacting via the screened effective potential generated by the other electrons and positive fermions in the gas. The exchange and correlation interactions amongst average electrons and positive fermions are treated in a mean field approximation. The two particular electrons must have a two-electron state antisymmetric with respect to their interchange, and correlation is taken into account by calculating multiple scattering beyond the Born approximation. One particular electron in the Fermi sea is considered the perturbation that induces the screened response of the average electrons and average positive fermions, resulting in the effective potential that the other particular electron feels.

With these assumptions, all of the interactions are known as a function of density $r_s$ and mass ratio *M/m*. For mass ratios *M/m* > 9, this simple model predicts that the system would have a charge density wave at the equilibrium density, invalidating the assumption of uniform density. Therefore I consider only mass ratios up to 9.

The wave function for two electrons must be antisymmetric in their exchange. For two antiparallel electrons, the spin wave function is antisymmetric and the spatial wave function is symmetric: the singlet state $V_S = V_0 - 3J$. For two parallel electrons, the spin wave function is symmetric and the spatial wave function is antisymmetric: the triplet state $V_T = V_0 + J$. Since $V_0 > 0$ and $J < 0$, the interaction is less repulsive for parallel electrons. BCS superconductivity considers pairing between antiparallel electrons.

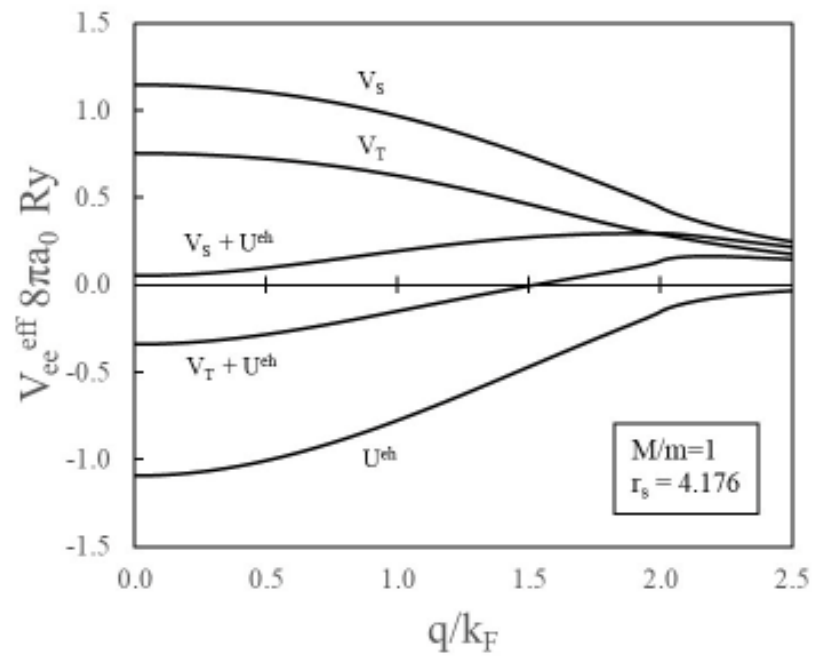

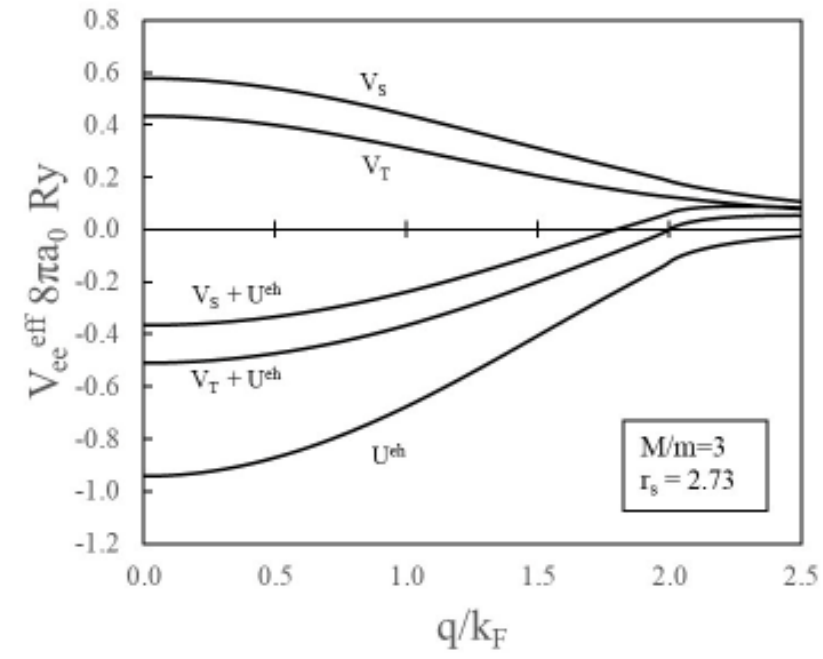

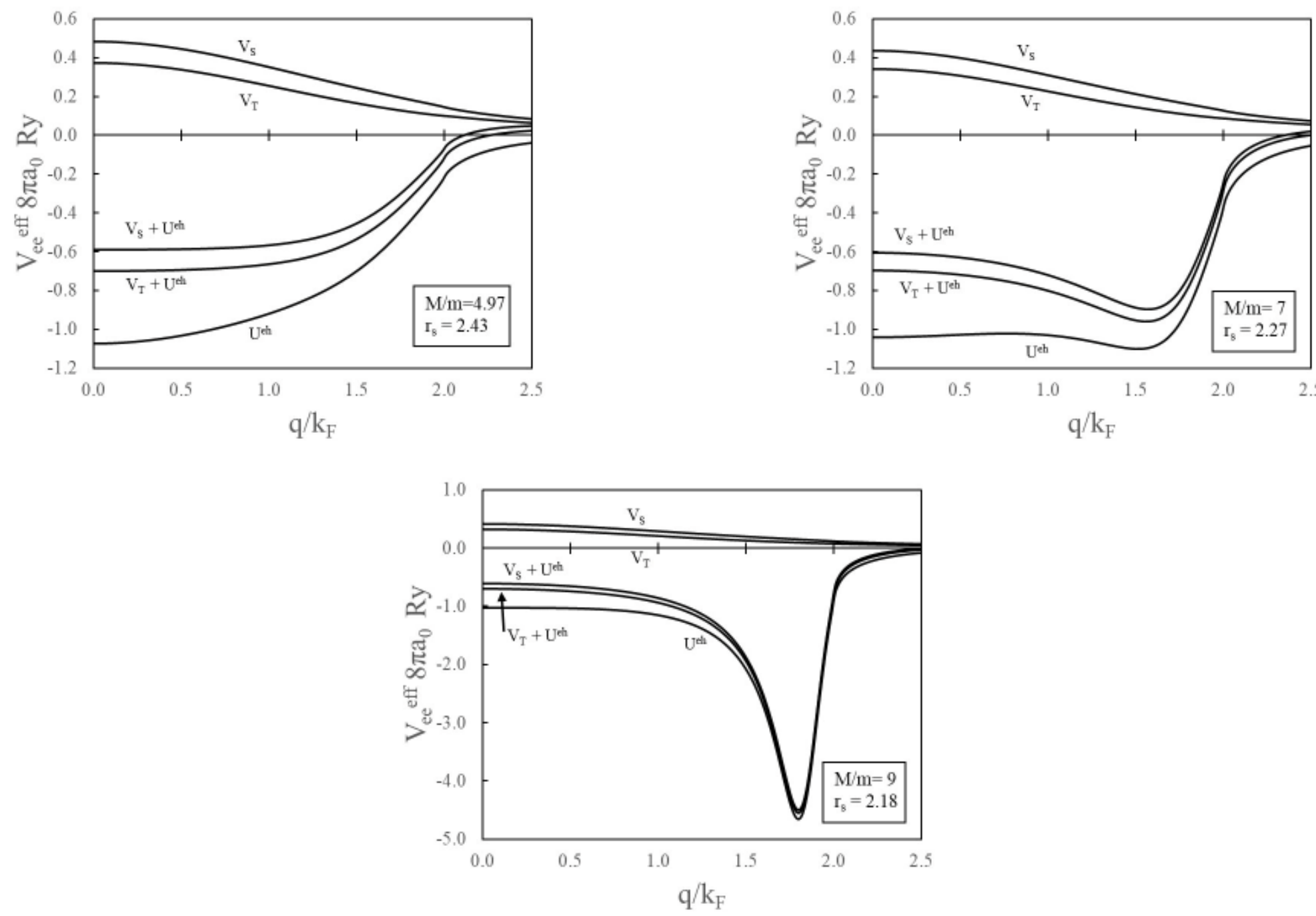

FIG. 8. The effective electron-electron interaction $V^{eeeee}$ using the current local field factors is plotted versus wave vector for different mass ratios M/m at the unique equilibrium density $r_s$ for each mass ratio. The repulsive portions $V_s$ and $V^T$ and the attractive term $U^{eh}$ are shown, along with the total singlet and triplet interactions. Rows show M/m = 1 and 3 (top), M/m = 4.97 and 7 (middle), and M/m = 9 (bottom).

Note that the repulsive terms $V_S$ and $V_T$ are the same as those for the electron gas in a uniform background and are independent of *M/m* (they depend only on $r_s$). They get smaller as the mass ratio increases because I am plotting at the equilibrium $r_s$ which decreases with mass ratio. The attractive term $U^{eh}$ is relatively constant at $q = 0$, because the equilibrium $r_s$ is closely 0.8 times the $r_s$ of the compressibility instability. The resulting net interaction at $q = 0$ gets more attractive for larger mass ratios because the repulsive terms are getting smaller.

At *M/m* = 1, the singlet state is always repulsive; the triplet state is slightly attractive for $q < 1.5k_F$, allowing possible triplet superconductivity at very low temperature. At *M/m* = 3, the singlet state is substantially attractive with minimum at $q = 0$. At *M/m* = 4.97 (triple point), the influence of the charge density wave becomes apparent. At *M/m* = 7, $U^{eh}$ has a minimum near $q = 1.55k_F$. At *M/m* = 9, the effective interaction has a deep minimum near $q = 1.8k_F$ which could favor p-wave or higher pairing. Linear response theory is clearly inadequate.

## VI. SUPERCONDUCTIVITY

The possibility of superconductivity in the electron-hole gas was investigated in Refs. [5, 8, 15] using an approximation and a numerical solution of the Eliashberg equation.

The Kukkonen-Overhauser [6] repulsive term has been incorporated into ab initio calculations of superconducting transition temperatures [16, 17] using superconducting density functional theory. This only covers the repulsive part of the interaction. For the electron-hole gas we are considering mass ratios in the range 1–9 and the Born-Oppenheimer approximation assumed in density functional theory is not applicable.

In conventional BCS superconductivity, the repulsive parameter μ is re-normalized into the Morel-Anderson pseudopotential μ* [19] because of the large difference in timescales between the electrons and the phonons. In this case, the mass ratios are close and this renormalization needs to be reconsidered. In the Eliashberg equation [18], the inputs are μ* and the electron-phonon spectral function $\alpha^2F(\omega)$. The Eliashberg equation is not suitable for the electron-hole gas when the masses are similar. Nevertheless, the previous investigators used this approach to estimate the superconducting transition temperature.

Because $V_{ee}^{eff}$ is very attractive, particularly near the charge density wave, I expect that superconductivity will be significantly enhanced. I hope this fact alone will stimulate new research into superconductivity in this system. Even though there are significant problems with this approach, I will follow Vignale and Singwi [5] and calculate the superconducting parameters λ and μ, which are simply averages over the Fermi surface of $V_{ee}^{eff}$ — simple integrals of $q \cdot V_{ee}^{eff}$ from $q = 0$ to $2k_F$. The repulsive term generates μ and the attractive term generates λ. The results are shown in Fig. 9.

The superconducting parameters λ and μ and their enhancement near the instabilities should be understood as precursors: they indicate that a strong pairing tendency exists in the normal phase, not that the divergent values are quantitatively reliable. A full treatment of the ordered phases would require methods that go beyond linear response theory.

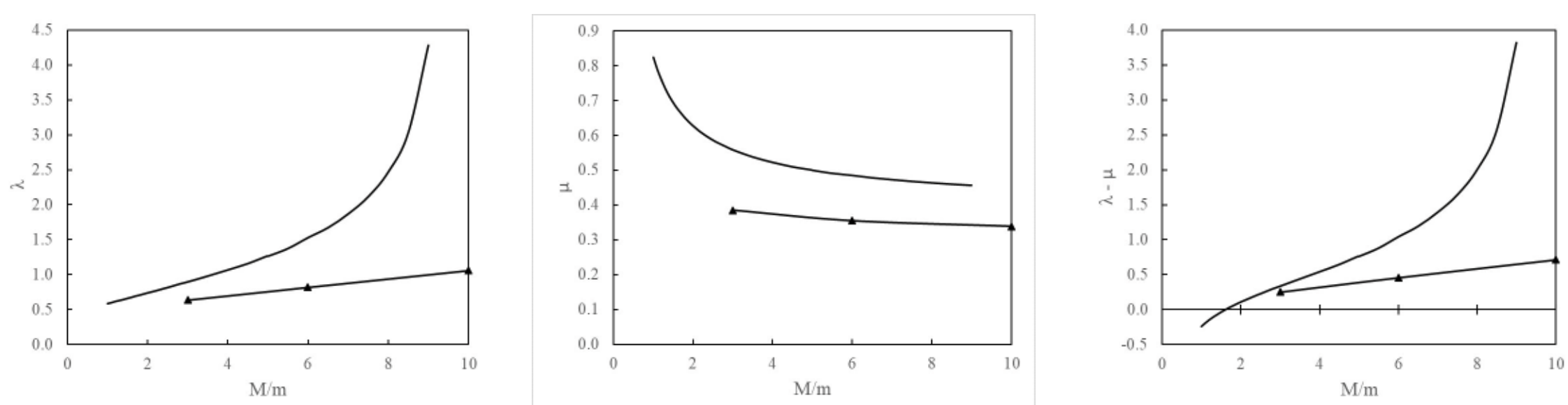

FIG. 9. The attractive superconducting parameter λ (left), the repulsive parameter μ (center), and the difference λ − μ (right) of the electron-positive fermion (hole) gas are plotted as a function of mass ratio M/m. The calculations are at the density $r_s$ of the unique energy minimum for each mass ratio. Also shown are previous results (triangles) from Vignale and Singwi [7] who used the Hubbard approximation for the local field factors.

As expected from Fig. 8, the attractive term λ is small for small mass ratios and gets large when the equilibrium density is near the density of the charge density wave, particularly at $M/m = 9$. The difference between the current results and those of Ref. [7] are due to their use of the Hubbard approximation. The repulsive term μ is independent of mass ratio and only depends on $r_s$. The net averaged interaction $-(\lambda - \mu)$ is attractive (negative) for $r_s > 1.5$.

The following estimate is purely illustrative. The physical conditions of this model — comparable electron and hole masses, no retardation separation, and no well-defined phonon spectrum — lie outside the regime for which the McMillan and Allen-Dynes formulae were derived. The numerical values of $T_c$ should not be taken as reliable quantitative predictions, but rather as indicators that the pairing tendency is strong enough to warrant a more rigorous calculation.

Following Ref. [7], I re-normalize μ using the Morel-Anderson [19] approach, assuming that the characteristic phonon frequency is now the Fermi temperature of the holes:

$$\mu^* = \mu / (1 + \mu \log(T_{Fe} / T_{Fh})) = \mu / (1 + \mu \log(M/m)) \tag{16}$$

The McMillan [20] and Allen-Dynes [21] formulae for the BCS superconducting transition temperature depend on μ*, λ, and a pre-factor that is an average phonon frequency. Following Ref. [7], I choose the Fermi temperature of the holes as the characteristic temperature pre-factor:

$$T_c = T_{Fh} \cdot \exp(-1.04(1 + \lambda) / (\lambda - \mu^*(1 + 0.62\lambda))) \tag{17}$$

$$T_{Fh} = 58200\ K / (M/m \cdot (r_s\, \varepsilon_B)^2) \tag{18}$$

where $\varepsilon_B$ is the background dielectric constant.

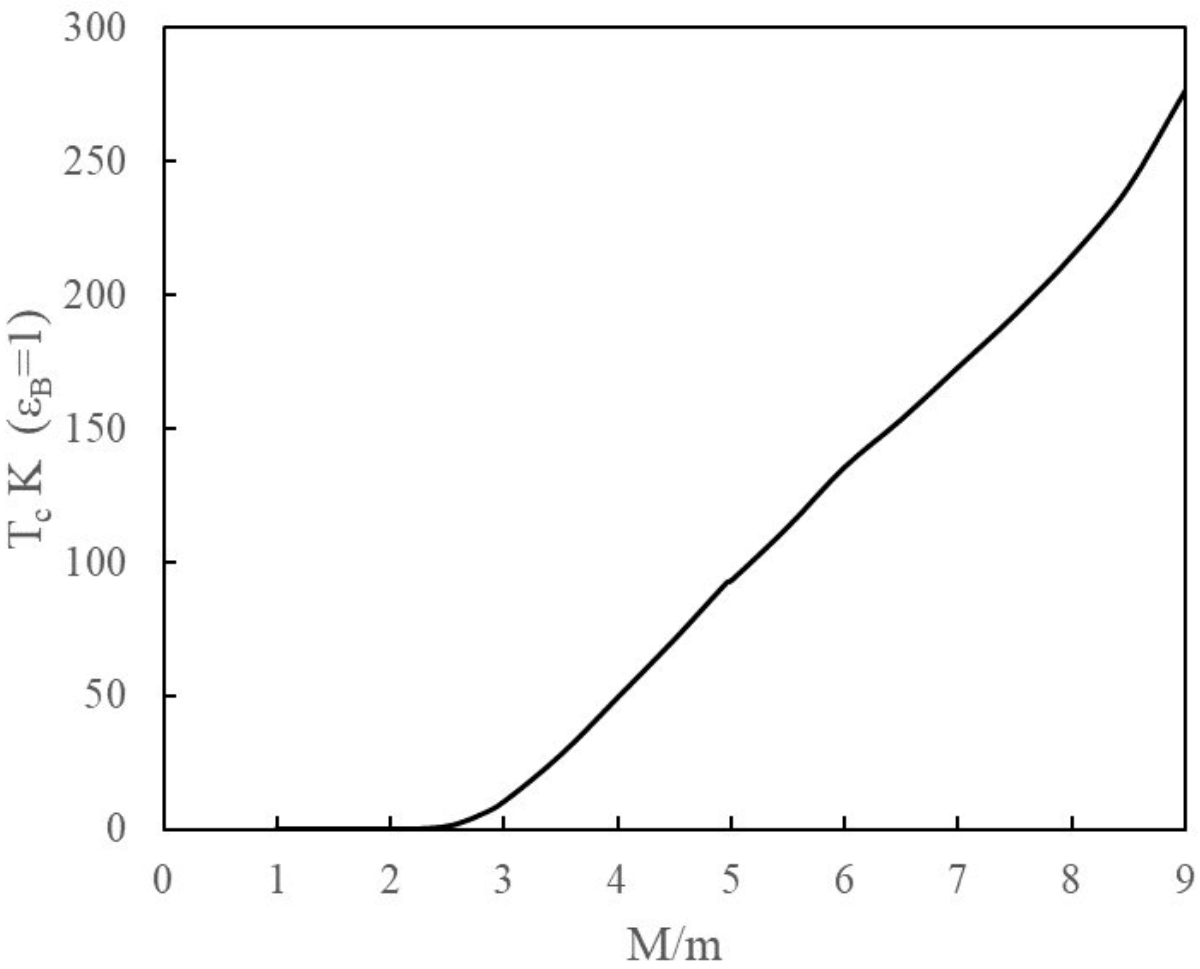

FIG. 10. Crude but indicative estimate of the superconducting transition temperature $T^c$ of the electron-hole gas as a function of mass ratio M/m. The calculations are at the unique equilibrium density $r_s$ at each mass ratio. If there were a background dielectric constant, the transition temperature is reduced by $1/\varepsilon^{l2}$.

With all of the caveats mentioned above, the superconducting transition temperature shown in Fig. 10 should be considered only as a crude indicative estimate. Nevertheless, quite high transition temperatures are predicted. The predicted transition temperature approaches room temperature, reaching approximately 275 K at M/m = 9. This figure alone should generate enough interest to consider the simple model seriously and to improve upon it or disprove it altogether. The source of the superconductivity is the additional screening provided by the holes.

Remember that this figure is generated at the unique equilibrium density for each mass ratio. At that equilibrium density there are no instabilities, but there is still substantial influence from the incipient instabilities that occur at lower densities. The fact that the superconducting transition temperature increases with mass ratio is obvious by simply looking at the effective electron-electron interaction in Fig. 8.

It should also be noted that the instabilities which enhance the superconducting pairing interaction may themselves preempt superconductivity. The $q$ = 0 instability likely corresponds to excitonic condensation for mass ratios below $M/m$ = 4.97, while the charge density wave instability at finite $q$ arrives first for $M/m$ > 4.97 and may similarly preempt or coexist with superconductivity. Whether Cooper pairing can emerge from within an existing CDW or excitonic phase is a fundamental open question that lies beyond the scope of this model.

The pre-factor in the equation for the transition temperature is the Fermi temperature for the holes which is inversely proportional to the mass ratio. If that were the only quantity that changed, the transition temperature would be reduced with increasing mass ratio. However the exponential factor increases faster than the inverse mass ratio which leads to the increasing superconducting transition temperature with mass ratio.

## VII. SUMMARY, CONCLUSIONS, AND CRITIQUE

The electron-positive fermion (hole) gas is a model that is simple to state, but a very difficult many-body problem. I have simplified this problem by making some clearly stated, but unverified and possibly controversial, approximations. With these approximations, it is possible to use known electron gas results to calculate the equilibrium density $r_s$, and the response functions as a function of wave vector for all mass ratios $M/m$. It is possible to calculate the shape of the ground state energy and take its first derivative to find the equilibrium density and the second derivative to find the point where the bulk modulus equals zero, which is the so-called compressibility instability. In addition, using the wave vector dependence of the local field factors of the electron gas, charge density waves are predicted for certain mass ratios and wave vectors. The electron-hole gas can have two instabilities and proximity to these instabilities strongly enhances electron-hole scattering and the effective electron-electron interaction for superconductivity calculations. Near the instabilities, the predicted $T^2$ electrical resistivity is strongly enhanced, as is the superconducting transition temperature.

I argue that these approximations are reasonable and have tested them where I can, but I cannot prove that they are correct. A major goal of this paper is to stimulate criticism and hopefully more research into this problem. Simply adding a second fermion charge carrier with its attendant screening opens up additional physics. A new parameter *M/m* — the mass ratio of the holes to electrons — in addition to density $r_s$ leads to surprising new results. I have provided the equations and the needed local field factors are readily available in Refs. [2, 12, 13]. The reader can easily reproduce all of the results of this paper using nothing more than Excel on a laptop as I did. The mathematics of the results are clear, but physical intuition is still lacking.

A basic prediction from the simple model is that if the mass ratio and density of the electron-hole gas are near an instability, the normal state electrical resistivity is due to electron-hole scattering, which would be large and varies as $T^2$. As the system is cooled, the resistivity would go down until the system becomes superconducting at an enhanced temperature. Application of pressure to the system would increase the density and reduce the size of the $T^2$ resistivity and depress the superconducting transition temperature.

The fundamental assumptions in the simple model are:

- The neutralizing background is uniform.
- The electron-hole correlation energy is constant with respect to volume and does not contribute to the pressure or bulk modulus.
- The local field factors from the uniform electron gas can be used and the contribution from electron-hole correlations is zero.
- The local field factors for the electron gas calculated at high density $r_s < 5$ can be extrapolated to low density $r_s > 20$.

The simple model of the electron-hole gas might stimulate the following research and extensions:

- Calculations of the ground state energy including the missing electron-hole correlation energy to establish the equilibrium density and the compressibility instability.
- Two-component quantum Monte Carlo calculations of the electron-hole response functions and thus local field factors.
- Develop an understanding and physical intuition about the behavior of the local field factors of the electron gas and electron-hole gas particularly at finite q.
- Electron-hole gas in a magnetic field.
- The nature of the new phases at the compressibility and CDW instabilities. Condensation, metal-insulator transition, exciton gas are possibilities at q = 0. Exotic phases near the CDW.
- Determine the competition between excitonic condensation, charge density wave order, and Cooper pairing. For $M/m < 4.97$ the compressibility instability at q = 0 likely represents excitonic condensation and would preempt superconductivity. For $M/m > 4.97$ the charge density wave instability at finite q arrives first and may similarly preempt superconductivity or coexist with it. Whether Cooper pairing can emerge from within an existing CDW or excitonic phase remains an open and unresolved question.
- p or d wave superconductivity near the charge density wave.
- Basic theory of superconductivity with a two-fermion gas.

- Generalize Morel-Anderson renormalization for the case of comparable electron and hole masses.
- Dynamical effects, frequency dependence, and retardation in the effective interactions.
- Go beyond linear response theory; consider renormalization.
- Simple extension to two dimensions.
- Multiple electron and hole pockets as done in Ref. [7] where the Fermi energies are not equal.
- Two different types of electrons with different masses and Coulomb and magnetic interactions.
- Thermal and electrical resistivity and the Wiedemann-Franz law.
- Second species could be a classical particle or boson.
- Two-component density functional theory and limits of the Born-Oppenheimer approximation.
- Mapping the simple model onto real systems.
- Pressure dependence as an experimental parameter.
- As the mass of the holes increases, the Fermi temperature decreases and the electrons remain very degenerate, but the holes may be similar to warm dense matter.

I have done very little new in this paper. The linear response theory of the electron gas and the electron-hole gas was done in the 1970s and 80s [4–8]. The results were expressed in terms of the local field factors. These were partially known and difficult to calculate back then. My contribution is to see that in the simple model, the now well-established local field factors could be used to calculate the phase diagram and the effective interactions in the electron-hole gas. The phase diagram agrees completely with that calculated using density functional theory in the local density approximation in Ref. [14] in their region of overlap.

However, seeing the effect of the compressibility instability and the charge density wave on the effective interactions and predictions of the electrical resistivity and superconducting transition temperature brings home the importance of the additional screening by the holes.

**Funding information**

The author received no external financial support for this research.

**Data availability**

All results presented in this paper are derived analytically from known local field factors of the electron gas. No numerical datasets were generated or analyzed. All calculations can be reproduced using the equations provided in Sections II and III.